\documentclass[%
preprint,
superscriptaddress,
 amsmath,amssymb,
 aps,
]{revtex4-1}

\usepackage{graphicx}
\usepackage{dcolumn}
\usepackage{bm}
\usepackage{amssymb}
\usepackage{caption}
\usepackage{subcaption}
\usepackage{amsmath}
\usepackage[T1]{fontenc}
\usepackage[utf8]{inputenc}

\begin{document}


\title{Inertial migration of a deformable capsule in an oscillatory flow in a microchannel}

\author{Ali Lafzi}
 \affiliation{Purdue University, West Lafayette, IN 47907, USA}
\author{Amir Hossein Raffiee}%
\affiliation{Purdue University, West Lafayette, IN 47907, USA}%

\author{Sadegh Dabiri}
 \affiliation{Purdue University, West Lafayette, IN 47907, USA}




\begin{abstract}
Dynamics of a deformable capsule in an oscillatory flow of a Newtonian fluid in a microchannel has been studied numerically. The effects of oscillation frequency, capsule deformability, and channel flow rate have been explored by simulating the capsule within a microchannel. In addition, the simulation captures the effect of the type of imposed pressure oscillations on the migration pattern of the capsule. An oscillatory channel flow enables the focusing of extremely small biological particles by eliminating the need to design impractically long channels. The presented results show that the equilibrium position of the capsule changes not only by the addition of an oscillatory component to the pressure gradient, but it also is influenced by the capsule deformability and channel flow rate. Furthermore, it has been shown that the amplitude of oscillation of capsules decreases as the channel flow rate and the rigidity of the capsule increases.
\end{abstract}

\maketitle


\section{\label{sec1}Introduction}
Manipulation of particles suspended in laminar flows in microchannels is commonly used in a variety of applications in bioparticle separation and filtration systems \cite{di2007continuous}. Depending on flow physics, the geometry of channel, and particles characteristics such as shape, size, and deformability, suspended particles focus in different regions of the channel, which can be used for sorting, isolating, or separating them for diagnostic and biological purposes \cite{gossett2010label, toner2005blood, gossett2012inertial, karimi2013hydrodynamic}. Diagnosing circulating tumor cells (CTCs) has always been of great interest, and due to heterogeneity of them, molecular assays and microfluidic technologies are required for cancer type-specific isolation \cite{van2011circulating}. Similarly, micro-scale vortices and inertial focusing were combined to extract CTCs from blood samples \cite{sollier2014size}. Another important advantage of these systems is being non-invasive; molecular cytogenetic techniques were used to identify fetal cells from maternal cells \cite{krabchi2001quantification}. 

One of the important shortcomings of current procedures, however, is their low flow throughput. Therefore, there have been many studies to address this issue in numerous applications. \citeauthor{di2008equilibrium} \cite{di2008equilibrium} used an asymmetrically curved channel to enhance the volume throughput and filter deformable particles with varying sizes.  Spiral microfluidic devices, having greater throughput compared to that of the existing microfluidic systems, have been introduced for cell separation and isolation \cite{lee2011high, warkiani2016ultra}. Removal of leukocytes (white blood cells) from the whole blood can be done with a continuous flow diffusive filter; isolation of plasma is also possible with simple modifications to the device \cite{sethu2006microfluidic}. \citeauthor{gossett2012hydrodynamic} \cite{gossett2012hydrodynamic} demonstrated an automated technology to assay cell deformability at very high throughputs. \citeauthor{ozkumur2013inertial} \cite{ozkumur2013inertial} have described a technology that can sort rare CTCs from the whole blood at a very high throughput of $10^7$ cells/s. Repetitive inertial focusing combined with micro-siphoning resulted in a microfluidic bioparticle concentrator \cite{martel2015continuous}. Stacked and cascaded inertial focusing strategies have been developed to achieve high throughput and concentrate desired particle size \cite{miller2016cascading}. Label and sheath-free inertial microfluidics were exploited for cell differentiation and blood fractionation with a high throughput \cite{hur2010sheathless, mutlu2017non}. 

In addition to inertial microfluidics, people have utilized electrical characteristics in the system. For instance, \citeauthor{gascoyne2004microfluidic} \cite{gascoyne2004microfluidic} have used dielectrophoretic field-flow-fractionation as a microfluidic approach to malaria detection. \citeauthor {vahey2008equilibrium} \cite{vahey2008equilibrium} implemented a continuous-flow and label-free microfluidic filter capable of separating cells based on their electrical polarizability.  Biochemical markers have also been used for cell sorting. \citeauthor{miltenyi1990high} \cite{miltenyi1990high} reported a simple magnetic system as a complement to flow cytometry. A micro-fabricated membrane was designed to perform in situ cell lysis with a very high efficiency \cite{zheng2007membrane}. These methods are, however, costly and complex and hence, less used in clinical applications compared to label-free approaches.

Although inertial microfluidics has a broad range of applications, most of the common practices are limited to control particles in the order of few microns in radius or larger (red blood cells, for instance) \cite{mutlu2018oscillatory}. This restriction is because small particles need to travel a very long distance to focus since the inertial lift force decreases for smaller particles. The theoretical length at which point particles travel to focus is inversely proportional to the cube of their radius. This can be shown by adopting a similar method \citeauthor{di2009inertial} \cite{di2009inertial} applied to finite particles. Hence for the case of nanoparticles, this traveling distance can reach up to the order of meters, which makes it practically impossible for design purposes \cite{mutlu2018oscillatory}. Nevertheless, it is vital to study small bioparticles, such as bacteria and fungi, due to their effective role in diagnosis. For instance, it is crucial to track microvesicles released by Glioblastoma tumor cells as they are beneficial in cancer patient care \cite{skog2008glioblastoma,  melo2015glypican}. There are only a few examples of working with smaller pathogens, such as separating pathogenic bacteria cells from diluted blood and detecting malaria parasites from blood \cite{mach2010continuous, warkiani2015malaria}. Moreover, a microchannel with multiple branches was shown to enable separation of smaller particles compared to that obtained with conventional pinched flow fractionation (PFF), but the size of separated particles is still not as small as desired \cite{takagi2005continuous}. Therefore, it is momentous to develop methods that can be applied to study small bioparticles and pathogens in a practical manner. 

In addition, the dynamics of capsules and particles in oscillatory microchannels has become of interest to many researchers recently; A simple equation containing only the Taylor deformation parameter and viscosity ratio has been proposed to estimate the threshold frequency for the capsule deformation behavior in an oscillatory shear flow in the Stokes flow regime \cite{matsunaga2015deformation}. \citeauthor{bryngelson2019non} \cite{bryngelson2019non} have done a non-modal and time-global Floquet stability analysis for a spherical capsule subject to a large-amplitude oscillatory extensional flow and found that all flow strengths (or a corresponding Weissenberg number) and oscillation frequencies are asymptotically stable despite experiencing some transient instabilities. The dynamics of formation and stability of particle pairs in an oscillatory flow in a microchannel has been investigated experimentally, and a linear correlation between the particle-particle interactions and flow velocity has been determined \cite{dietsche2019dynamic}.

In this work, we implement the idea of oscillatory inertial microfluidics in microchannels, expressed by \citeauthor{mutlu2018oscillatory} \cite{mutlu2018oscillatory}. The advantage of this method is that by changing the direction of the flow at a certain frequency, the virtual length at which the particle can travel is extended beyond the physical length of the channel. Due to the symmetrical flow conditions, the directions of forces acting on the particle in the wall-normal direction remain the same. Therefore, small particles corresponding to small values of particle Reynolds number (${Re_p}<0.1$) can reach their focal position in a short physical length of the device, which was otherwise unfeasible. Furthermore, it will be shown that the frequency of flow oscillation influences the equilibrium position of capsules in the channel. This parameter can be used as an extra tool for direct control over the dynamics of capsules.

\section{\label{sec:level1}Methodology}
\begin{figure}[h]
\centering
  \begin{subfigure}[h]{.485\textwidth}
  \centering
  \includegraphics[height=.345\textheight]{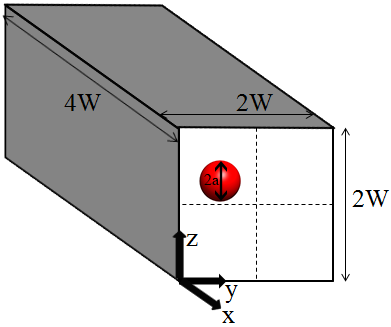}
  \caption{}
  \label{geometry}
  \end{subfigure}
  ~
  \begin{subfigure}[h]{.485\textwidth}
  \centering
  \includegraphics[height=.345\textheight]{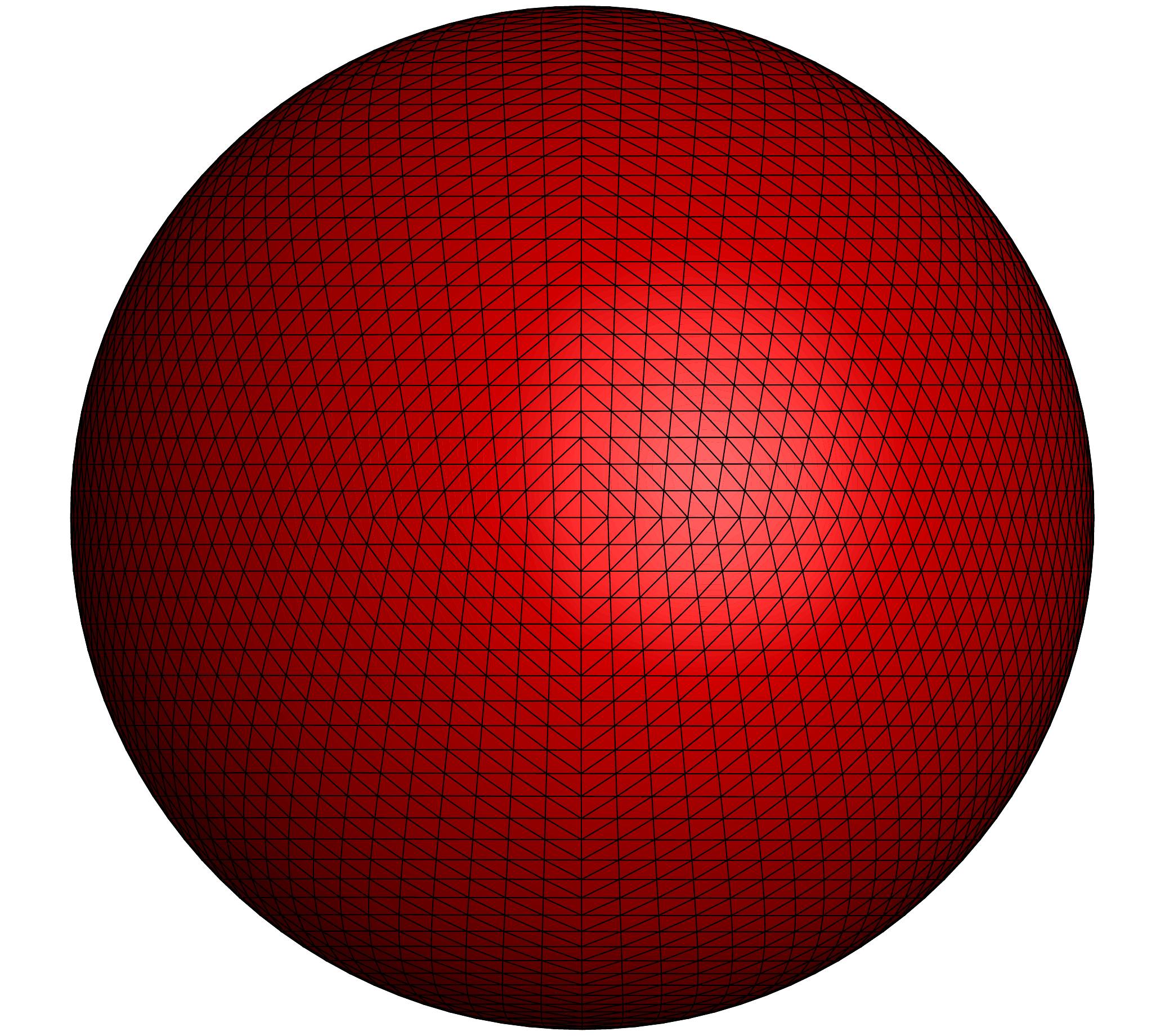}
  \caption{}
  \label{discretization}
  \end{subfigure}
  ~
   \caption{(a) Schematic of the problem setup and (b) An example of the capsule discretization}
     \label{setup}
\end{figure}

A single deformable capsule has been placed in a laminar flow of an incompressible Newtonian fluid in a microchannel with a square cross-section. A schematic of the configuration is illustrated in Fig. \ref{geometry}. The density and viscosity of the inner Newtonian fluid inside the capsule are the same as those of the outer one. The front-tracking method \cite{unverdi1992front} is used to track the position of the interface. The front consists of Lagrangian grid points connected by triangular elements (Fig. \ref{discretization}). In this method, the main governing equations (equations \eqref{eq:1} and \eqref{eq:2}) are solved for both fluids inside and outside of the capsule on a fixed Eulerian grid. The local velocity of the fluid is used to move the Lagrangian points on the capsule membrane under the assumption of no-slip condition on the capsule membrane. The governing equations to be solved in the entire computational domain are the following:
\begin{equation}\label{eq:1}
\nabla \cdot \textbf{u}=0,
\end{equation}
\begin{equation}\label{eq:2}
\frac{\partial (\rho \textbf{u})}{\partial t}+\nabla \cdot (\rho \textbf{uu})=-\nabla P+\nabla \cdot[\mu(\nabla \textbf{u}+\nabla \textbf{u}^T)] +\textbf{F},
\end{equation}
where ${\rho}$ and ${\mu}$ are the density and viscosity of both fluids respectively, $P$ represents the pressure, \textbf{u} is the velocity vector, $t$ is the time, and $\textbf{F}(\mbox{\boldmath$x$},t)=\int_{\partial B}^{} $\mbox{\boldmath$f$}($\mbox{\boldmath$x_i$},t)\delta($\mbox{\boldmath$x$}-$\mbox{\boldmath$x_i$}) dV$, which is the smoothed representation of the membrane elastic force. This is shown by placing a Dirac delta function within the integral, where $\mbox{\boldmath$x$}$ is an arbitrary location in the whole computational domain and $\mbox{\boldmath$x_i$}$ is such position on the capsule membrane. The given delta function is defined as:
\begin{equation}\label{eq:9}
\delta(X)=\tilde{D}(x)\tilde{D}(y)\tilde{D}(z),
\end{equation}
\begin{equation}
  \tilde{D}(x) =
    \begin{cases}
     \frac{1}{4\Delta}(1+\cos(\frac{\pi}{2\Delta}(x))) &  |x|\le 2\Delta,\\
      0 & \text{otherwise,}
    \end{cases}       
\end{equation}
where $\Delta$ is the constant Eulerian grid size. The capsule membrane is assumed to be an infinitely thin sheet of elastic material. Using the Skalak model \cite{skalak1973strain}, we assign the following strain energy function to the capsule membrane:
\begin{equation}\label{eq:3}
\textit{W} =\frac{E_s}{12}((\epsilon_1^2+\epsilon_2^2-2)^2+2(\epsilon_1^2+\epsilon_2^2-2)-2(\epsilon_1^2\epsilon_2^2-1))+\frac{E_a}{12}(\epsilon_1^2\epsilon_2^2-1)^2,
\end{equation}
Here, $\epsilon_1$ and $\epsilon_2$ are the principal strains, and $E_s$ and $E_a$ represent the shear and area dilatation moduli, respectively. We consider $\frac{E_a}{E_s}=2$ according to \citeauthor{kruger2014interplay} \cite{kruger2014interplay}. Bending resistance has not been taken into account as its effect is negligible compared to that of the shear modulus \cite{parker1999deformation}. A finite element method \cite{charrier1989free} is used to calculate the forcing term $\mbox{\boldmath$f$}$. The membrane surface is discretized with a large number of triangular elements so that they remain approximately flat in case of undergoing large deformations. The resultant elastic force ($\mbox{\boldmath$f$}(\mbox{\boldmath$x_i$},t)$) acting on the membrane is found using the principle of virtual work. The validation of this model against previously published results and a more detailed explanation of this whole methodology can be found in \citeauthor{raffiee2017deformation} \cite{raffiee2017deformation}. 

Two forms of oscillatory pressure gradients have been applied along the channel ($x$ direction) to change the direction of the flow symmetrically: a) A cosine wave with a constant amplitude (in the form of ${P_0}cos({\omega}t)$) b) A square wave with the same average and $\omega$ as that of the cosine wave; so $P_0$ is different. The periodic boundary condition is applied in the $x$ direction, and the no-slip condition is applied on the walls in the $y$ and $z$ directions. $W$, $U_c$ (maximum velocity of the steady flow), and $\frac{W}{U_c}$ are used to nondimensionalize all lengths, velocities, and time, respectively. In other words, $x^*={\frac{x}{W}}$, $u^*={\frac{u}{U_c}}$, $t^*={\frac{t}{{\frac{W}{U_c}}}}$, $P^*={\frac{P}{\mu{\frac{U_c}{W}}}}$, $T^*={\frac{T}{{\frac{W}{U_c}}}}$ (where $T$ is the period), and $\omega^*={\frac{2\pi}{T^*}}$. Two dimensionless parameters describe the motion of the capsule: (i) Reynolds number, $Re={\frac{{\rho}U_c2W}{\mu}}$, expressing the ratio between inertial forces to viscous ones (ii) Laplace number, $La={\frac{2aE_s{\rho}}{\mu^2}}$, which denotes the deformability of the capsule, where low $La$ corresponds to highly deformable capsules, and high $La$ represents more rigid particles. The blockage ratio of the capsule (${\frac{a}{W}}$) is constant and equal to $0.3$. The capsule is assumed to have a spherical initial shape and is released at ${\frac{y}{W}}=0.55$ and ${\frac{z}{W}}=1.07$. The initial location of the capsule is arbitrary because it only affects the initial, transient stage of the capsule migration and does not influence its long-term behavior and focusing point \cite{shin2012dynamics, kim2015inertial, lan2013lateral, villone2019lateral}. The axes of symmetry are also avoided for the initial location of the capsule. The time step of the simulation is restricted by the Courant-Friedrichs-Lewy (CFL) number, which is set to $0.9$. An Eulerian grid of $200\times 115 \times 115$ in the $x$, $y$, and $z$ directions, respectively, and 48672 triangular elements for the discretization of the capsule surface are used in most of the simulations. For the cases with $Re>10$, we use $256\times 152 \times 152$ Eulerian grid points and 80000 Lagrangian elements.

\section{\label{sec:level1}Results and Discussion}
\subsection{\label{sec:level2}Sinusoidal oscillatory flow}

We study the effects of inertia, deformability, and pressure oscillations frequency on the migration of the capsule by adjusting $Re$, $La$, and $\omega^*$. $Re$ ranges between $5$ and $37.8$, $La$ ranges between $1$ and $500$, and  $\omega^*$ values are chosen such that for a channel with a square cross-section of $100 \mu m$ and water as the working fluid at room temperature, the frequency values range between 2Hz and 200Hz, which is mostly reported in the literature \cite{dincau2020pulsatile}. For each case, the capsule is placed at the same initial location in the channel and is tracked until it reaches its focal position. This equilibrium position is a result of the competition between the forces acting on the capsule, namely the wall effect and the capsule deformation-induced lift forces, both acting towards the channel center, and the shear gradient force acting towards the wall \cite{fay2016cellular, zhang2016fundamentals}. Magnus and Saffman lift forces are often very small and negligible compared to the other mentioned components \cite{zhang2016fundamentals}. The role of the boundary wall is to generally retard the particle motion. When the particle moves parallel to the wall, it experiences a transverse lift force that repels it away from the wall \cite{zhang2016fundamentals}. The existing curvature of the parabolic fluid velocity profile makes the magnitude of the relative velocity of the fluid to that of the particle on the wall side much higher than the channel center side.  This dissymmetry causes a low pressure on the wall side resulting in a shear gradient lift force that pushes the particle towards the closer wall from the channel center \cite{zhang2016fundamentals}. To the best of our knowledge, there is no quantitative expression for the shear gradient force in the literature to date. Following the analytical results of \citeauthor{chan1979motion} \cite{chan1979motion}, the deformability-induced lift force for droplets or bubbles that have a distance higher than their diameter from the wall, which is the case in our simulations, is given by \cite{stan2013magnitude}:
\begin{equation}\label{deformation force}
F_{L,deformation} =-75.4Ca_p\mu V_{avg}a(\frac{a}{W})^2\frac{d}{W},
\end{equation}
\begin{equation}\label{capillary}
Ca_p={\frac{\mu V_{avg}}{\gamma}}{\frac{a}{W},}
\end{equation}
Where $Ca_p$ is the particle capillary number, $V_{avg}$ is the average velocity of the carrier fluid across the channel, $d$ is the distance of the particle from the channel center, and $\gamma$ is the surface tension at the interface. Here, $f(\lambda)$ in the original equation, (where $\lambda$ is the viscosity ratio between the inner and outer fluids), is replaced with its value of $-75.4$ evaluated at $\lambda=1$. The negative sign indicates that this force acts towards the center of the channel.

\begin{figure}
\centering
  \begin{subfigure}[t]{.485\textwidth}
  \centering
  \includegraphics[height=.345\textheight]{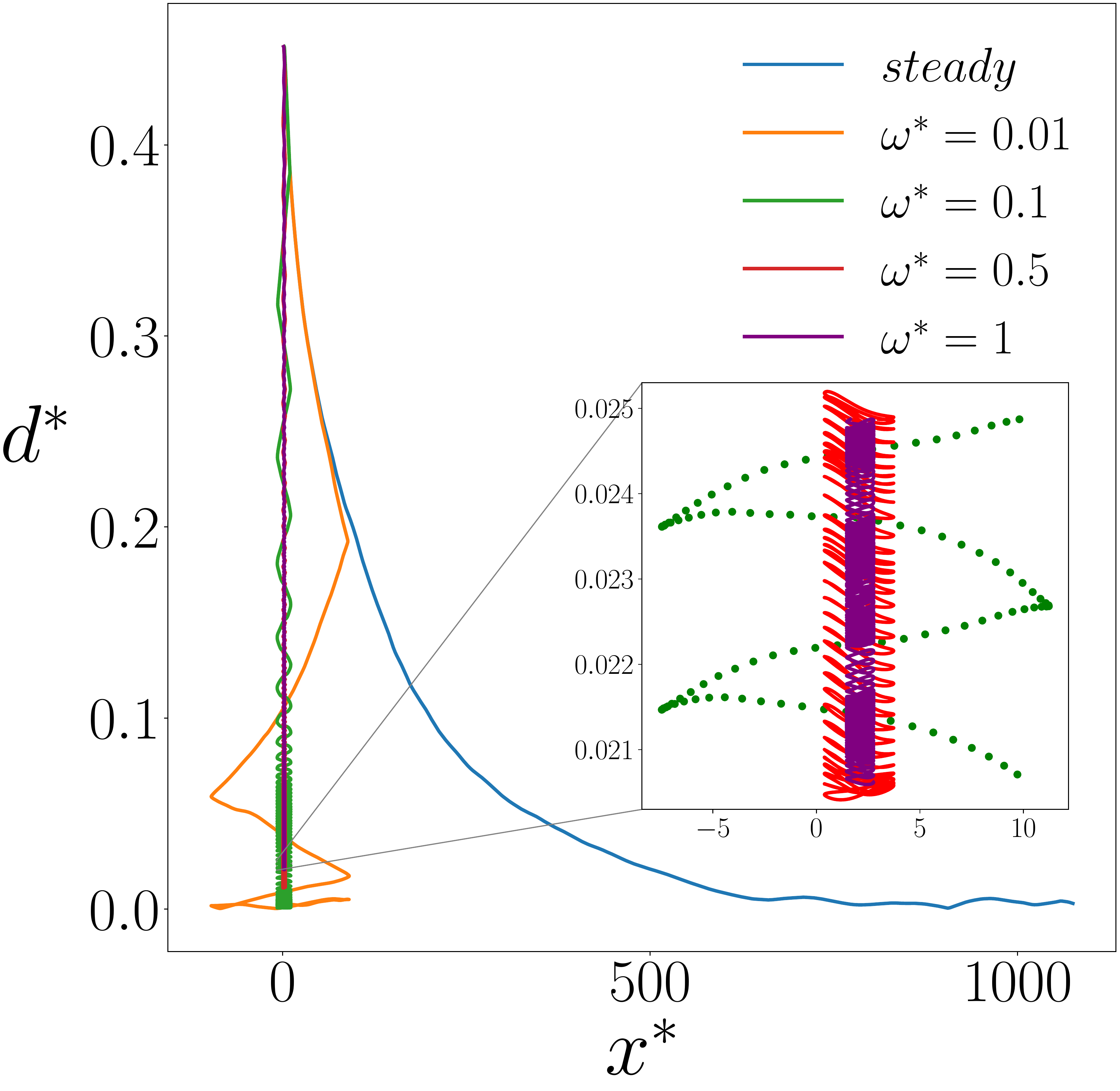}
  \caption{}
  \label{position}
  \end{subfigure}
  ~
  \begin{subfigure}[t]{.485\textwidth}
  \centering
  \includegraphics[height=.345\textheight]{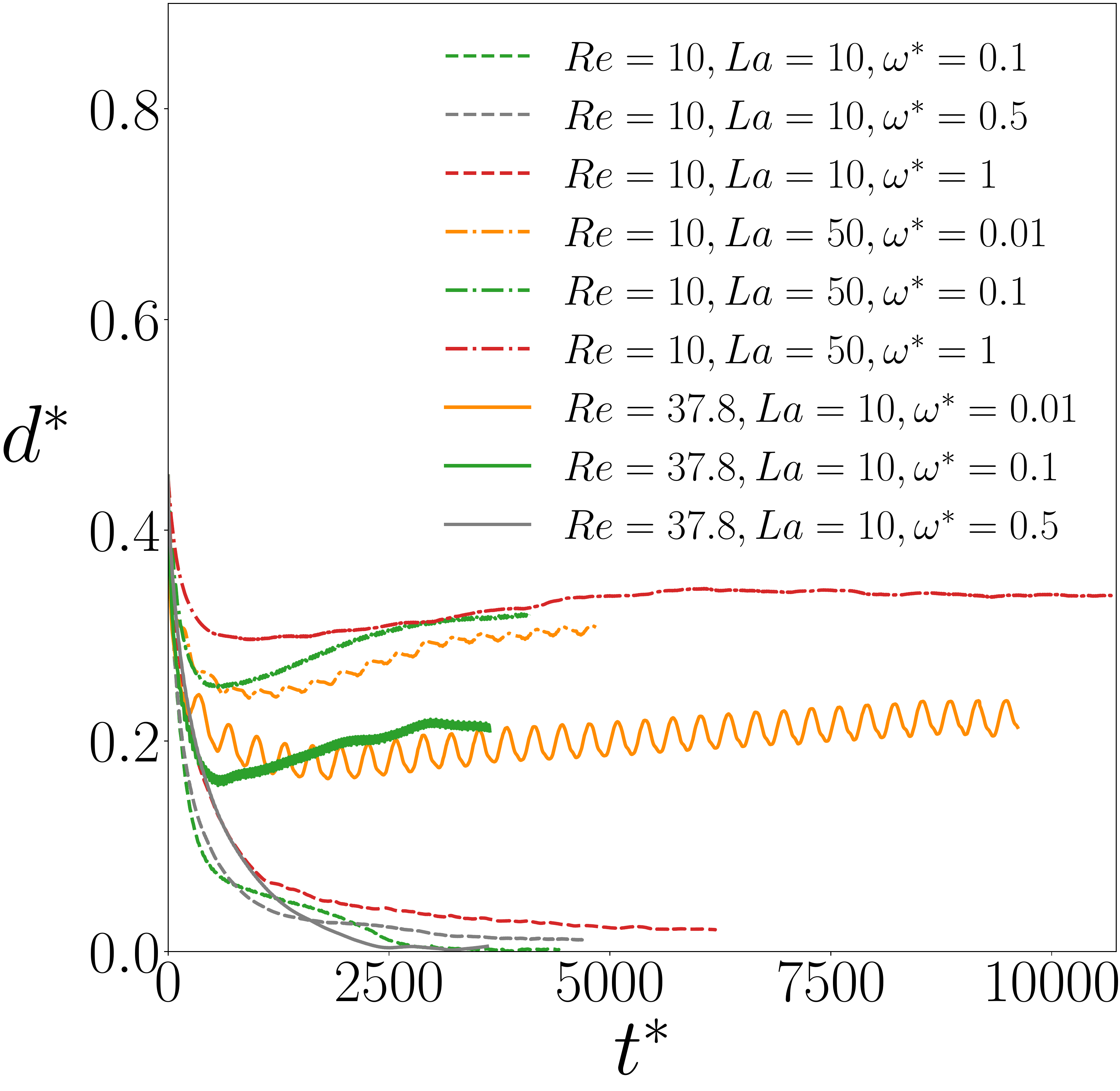}
  \caption{}
  \label{time}
  \end{subfigure}
  ~
  \begin{subfigure}[t]{.485\textwidth}
  \centering
  \includegraphics[height=.345\textheight]{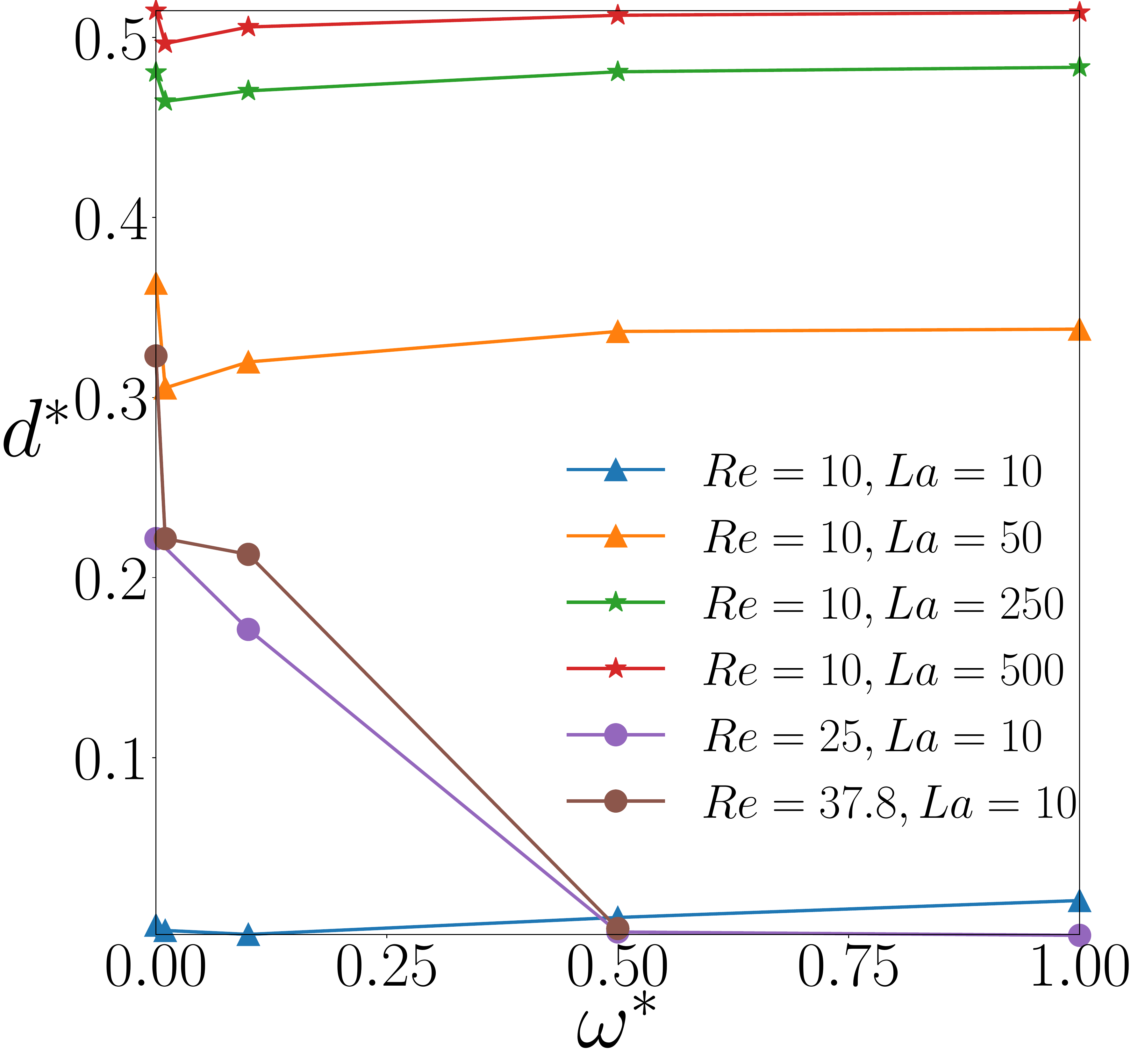}
  \caption{}
  \label{frequency}
  \end{subfigure}
  ~
  \begin{subfigure}[t]{.485\textwidth}
  \centering
  \includegraphics[height=.345\textheight]{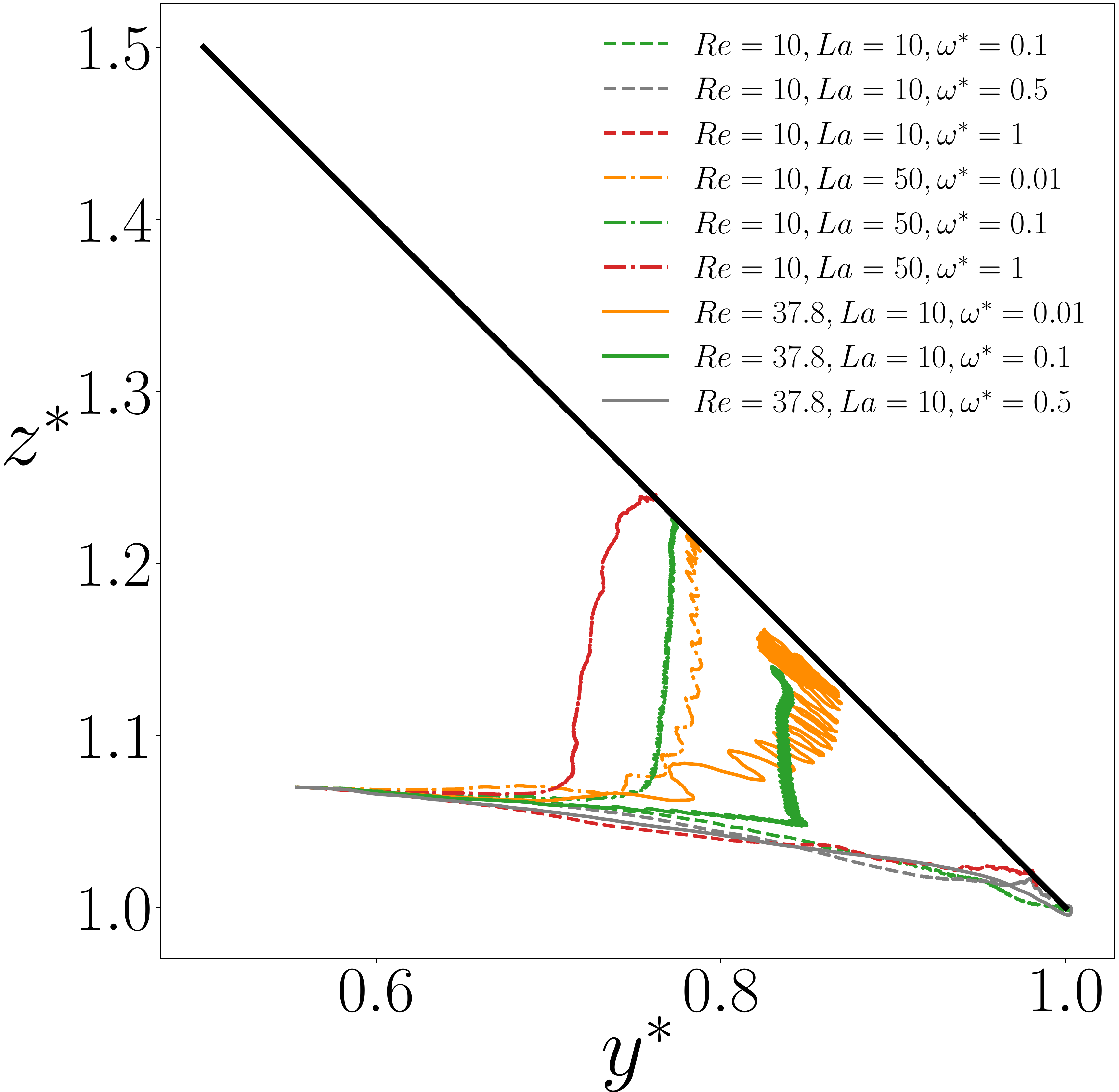}
  \caption{}
  \label{lateral}
  \end{subfigure}
   \caption{Evolution of the capsule motion for different cases by illustrating the time dependent progress in its trajectory (a) versus the flow direction for $Re=10$ and $La=10$, (b) versus time, (c) the equilibrium position versus the frequency, and (d) on the channel cross section with the same legend as that of the (b)}
     \label{different frequency}
\end{figure}

Fig. \ref{different frequency} depicts the effect of frequency, capsule deformability, and channel flow rate on the capsule dynamics. In these sub-figures, $d^*$ is the dimensionless distance of the capsule from the channel center. Fig. \ref{position} illustrates this distance versus the location of the capsule along the channel. It shows that the capsule has to travel a very long distance along the flow direction in the steady flow to reach its equilibrium position. On the other hand, increasing the frequency lowers the amplitude of capsule motion due to the frequent change in the flow direction, which means that the capsule travels a shorter distance to focus; this implies that by increasing the frequency of the pressure gradient wave, the necessary length of the microfluidic channel can be decreased significantly. However, increasing the frequency steadily has its own limitations, as it is challenging to create very high frequencies in practice \cite{vishwanathan2020generation}. In particular, the case of $\omega^*=0.1$ is plotted by dots at equal time intervals. The varying distance between its consecutive points is due to the sinusoidal pressure gradient.

Fig. \ref{time} shows the time evolution of the distance of the capsule from the center of the channel for several cases. The most important feature of this figure is the change in the focal point of the capsule due to the change in the frequency of oscillation. This is because the wall effect and shear gradient induced lift forces directly depend on the flow velocity, which has a different quantitative profile for different frequency values. This, in turn, changes the lateral location of the capsule leading to changes in the deformation-induced lift force \cite{zhang2016fundamentals}.

Fig. \ref{frequency} shows the equilibrium location of all the cases in this study. At $Re=10$ and all the $La$ numbers, there is an intermediate value of frequency at which the capsule focuses closer to the channel center compared to the steady and other oscillatory cases. This frequency depends on both the capsule deformability and channel flow rate. Since $V_{avg}$ is a function of time in the oscillatory flows, an appropriate approach to interpret it is to refer to its average in a portion of the corresponding periodic cycle when the flow is in a single direction (half a period). For the higher frequency at the same amplitude of the pressure gradient, the absolute value of the average of $V_{avg}$ in each half a periodic cycle is lower. This is confirmed by comparing the corresponding flow rates in the channel. This further implies that the maximum of $U_c$ in each cycle decreases by increasing the frequency. Therefore, at higher frequencies, both deformation and shear gradient forces are lower on average; this is because of equation \eqref{deformation force} and the resultant lower difference between the relative velocities on the wall and channel sides, respectively. From Fig. \ref{frequency}, we can conclude that by increasing the frequency to a particular value, the decrease in the shear gradient force is more than that of the deformation force, making the latter the dominant one. Above this critical frequency, the decrease in the deformation force is more, which makes the shear gradient lift the overcoming force. At $Re=25$ and $Re=37.8$, however, increasing the frequency leads to a focusing point closer to the center. This means that the average deformation force always overcomes the average shear gradient lift at higher Reynolds. The wall-induced lift force for all the studied cases here is negligible as the capsule paths are far from the wall. \cite{zhang2016fundamentals} 

It is also notable from Fig. \ref{frequency} that at the fixed $Re=10$, there is an intermediate value of $La$ number ($La=50$ in the figure) for which the differences between the capsule focal points in the steady and oscillatory flows at the depicted frequencies are higher than those for the other capsule deformabilities. This is because by changing the frequency, the capsule generally begins to travel a different trajectory due to the change in the flow motion pattern and values of the shear gradient lift force. Since the deformation lift force also depends on the capsule location and shear, its magnitude will also be different by changing the frequency. This means that the change in the values of both lift forces contributes to different focal points at different frequencies. However, when $La$ is very high ($La=500$ and $La=250$ for instance), the capsule cannot deform much resulting in close values of the deformation-induced force. This behavior can also be observed in equation \eqref{deformation force} as the $Ca_p$ is a small value due to the consequence of the $Ca_p=\frac{Re_p}{La}$ relationship \cite{schaaf2017inertial}. Therefore, the difference between the focal points at different frequencies only comes from different magnitudes of the shear gradient force. On the other hand, at $Re=10$ and $La=10$, we can observe that for all the studied frequencies, the capsule focuses close to the center. This is because $La$ is very low for this case, and so the deformation force dominates the shear gradient force repelling the capsule towards the center regardless of the flow motion pattern. Thus, the difference between the focal points at different frequencies in this case only comes from different magnitudes of the deformation force. 

Furthermore, increasing $Re$ increases the distance between the focal points of the capsule in the flows with the frequencies depicted in Fig. \ref{frequency} (including the steady flows) at the same $La$. This is due to the higher change in the corresponding shear gradient and deformation lift forces as a consequence of the higher difference between the absolute values of the average of $V_{avg}$ in each half a periodic cycle at different frequencies. This is also confirmed by comparing the differences among the average of flow rates in the steady and oscillatory flows by keeping $\omega^*$ and $La$ fixed and changing $Re$ only. Another possible explanation for this phenomenon could be the effect of the $Re_p$ term in the inertial lift force equation reported in \citeauthor{di2009particle} \cite{di2009particle} ($F_{L,inertial} =-10Re_p\mu V_{avg}a\frac{d}{W}$) although the restricted criteria of $\frac{d}{W}<0.2$ for validity of this equation does not apply to all the cases in our work. 

 Due to the square shape of the channel cross-section, the inherent symmetry is reduced compared to the Segre and Silberberg effect \cite{segre1961radial}, which is the appearance of colloidal particles on the circular annulus. Therefore, there is a set of discrete equilibrium points on the main axes and diagonal of the cross-section instead \cite{prohm2014feedback}. It has been reported in the literature that larger particles migrate towards the diagonal, where they are further away from the walls, while smaller particles migrate towards the main axes \cite{schaaf2017inertial}. Fig. \ref{lateral} shows the lateral migration of the capsule on the channel cross-section, where the solid black line is the diagonal. While most of the cases including all the steady ones focus on the diagonal, there are a few oscillatory cases that have reached their equilibrium location not exactly on the diagonal but rather close to it, which is also observed in other works \cite{schaaf2017inertial, villone2019lateral}. As capsules become stiffer, they tend to focus farther from the channel center \cite{raffiee2017elasto, kilimnik2011inertial}. Moreover, it is apparent from this figure that the $y$ and $z$ positions of the capsule are also oscillatory in addition to the $x$ position as shown in Fig. \ref{position}. Therefore, all the equilibrium points reported in Fig. \ref{frequency} are the average values in each corresponding oscillatory cycle of the trajectory.



It is important to note that we have shown that the numerical results are independent of the distance between $2$ consecutive capsules in an infinite domain in the X-direction. This has been done by comparing the distance of the capsule from the center at $Re=10$, $La=50$, and $\omega^*=0.1$ for three different distances of $4W$, $6W$, and $8W$ between $2$ consecutive capsules in the X-direction. The maximum difference between the capsule distances from the center for $L=4W$ and $L=6W$ is $0.65\%$ or $0.002W$, and the one between those of $L=4W$ and $L=8W$ is $0.77\%$ or $0.002W$. The results have also been shown to be mesh independent, by comparing the focal positions for the case of $Re=10$, $La=50$, and $\omega^*=0.1$ with two different grids of $200\times 114 \times 114$ and $256\times 152 \times 152$. The difference between their focusing distances from the center is $2.66\%$ or $0.0085W$. To validate our numerical results, the focusing distance as a function of $La$ has been compared to those of a previously published work \cite{schaaf2017inertial}. The maximum error observed in the capsule trajectory at different $La$ numbers is $1.15 \%$. This validation has also been done in a previous work \cite{raffiee2017elasto}.

\begin{figure}[h]
  \centering
  \includegraphics[width=3.3in]{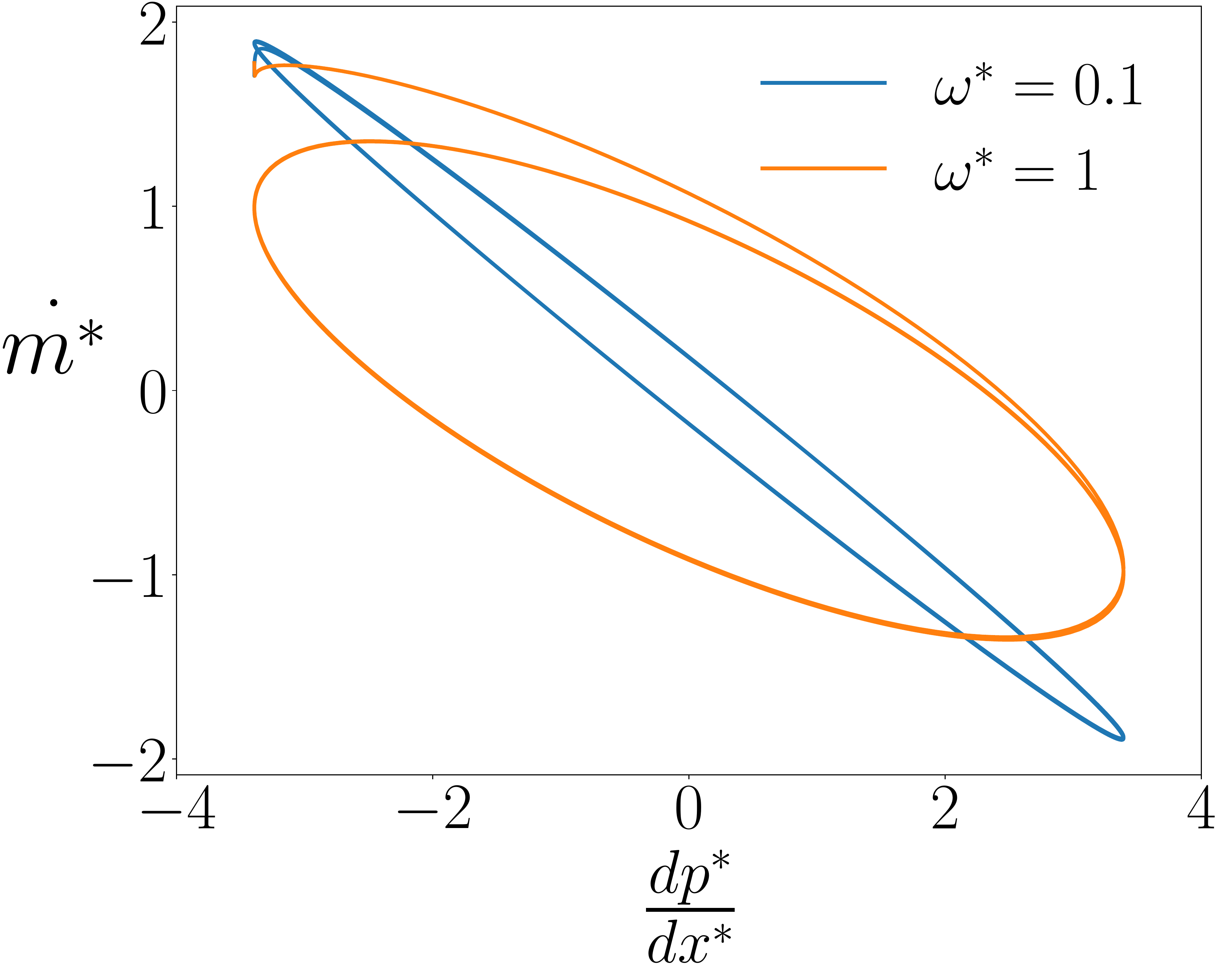}
   \caption{Hysteresis loop at two different frequencies for $La=10$ and $Re=10$}
   \label{hysteresis}
\end{figure}

The introduced oscillatory flow is not quasi-steady since there is a phase lag between the imposed pressure gradient and the resultant flow parameters such as the velocity field; this is illustrated in Fig. \ref{hysteresis} by plotting the mass flow rate against the pressure gradient forming a hysteresis loop. Duo to the fluid inertia, there is a delay for the flow in response to the external oscillatory driving force. This effect is more significant for higher values of frequency \cite{noguchi2010dynamic}. As the frequency decreases, the region becomes narrower and eventually approaches a line (quasi-steady). It can also be concluded that the more the hysteresis is, the lower is the required device length.

\begin{figure}[h]
  \centering
  \includegraphics[width=3.3in]{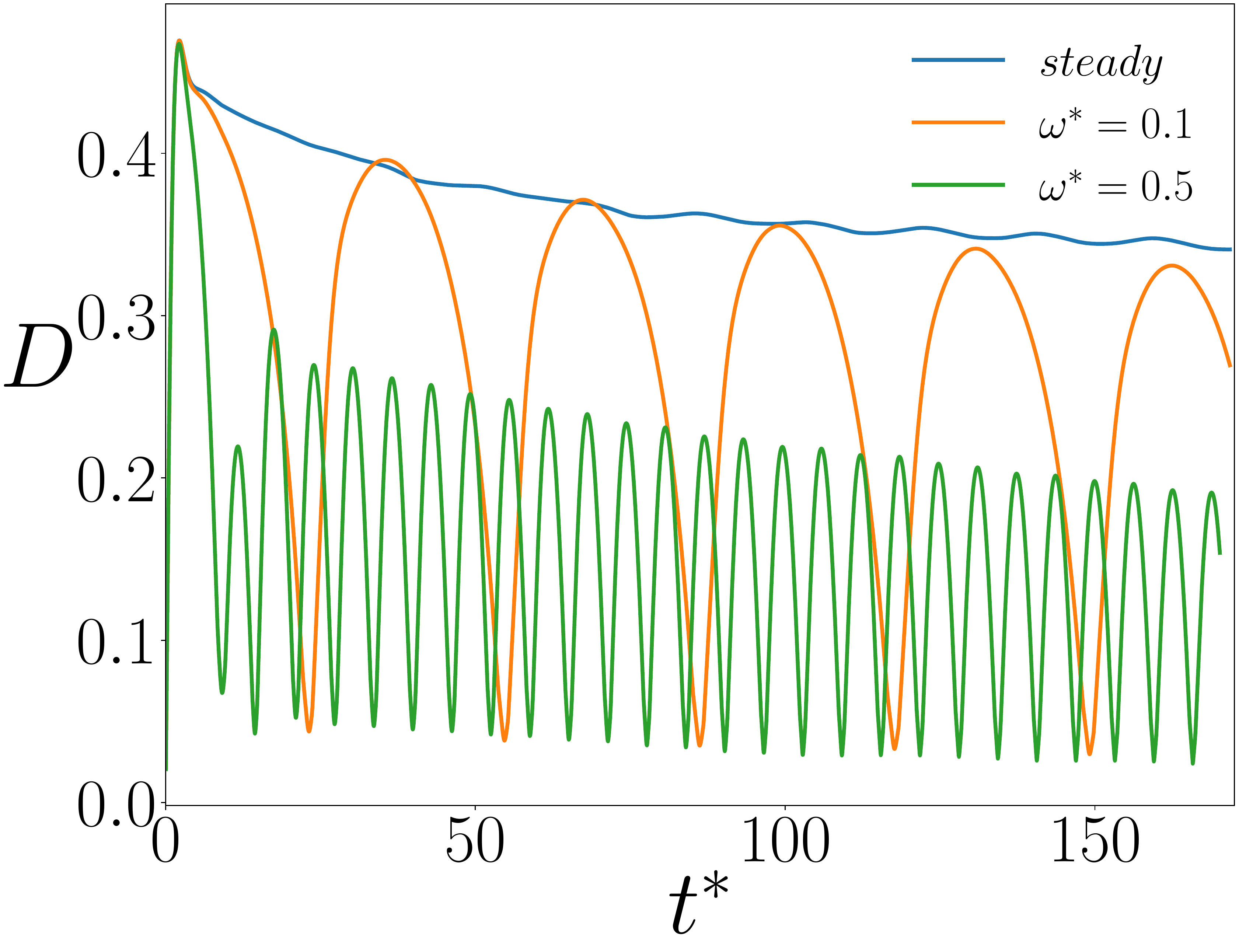}
   \caption{Deformation of the capsule at three different frequencies for {$La=10$} and $Re=37.8$}
   \label{deformation}
\end{figure}

\begin{figure*}
  \begin{subfigure}[b]{0.3\columnwidth}
    \includegraphics[width=\linewidth]{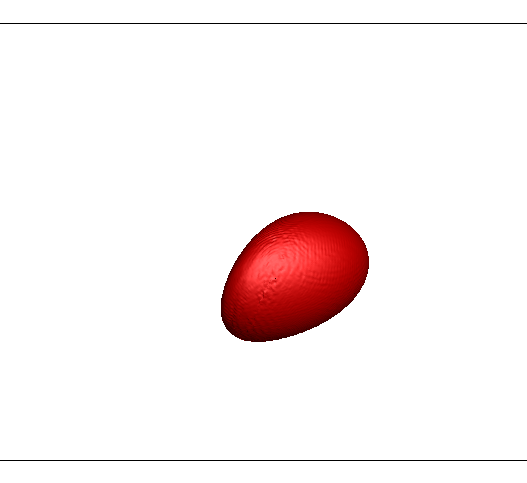}
    \caption{}
    \label{left to right}
  \end{subfigure}
  \begin{subfigure}[b]{0.3\columnwidth}
    \includegraphics[width=\linewidth]{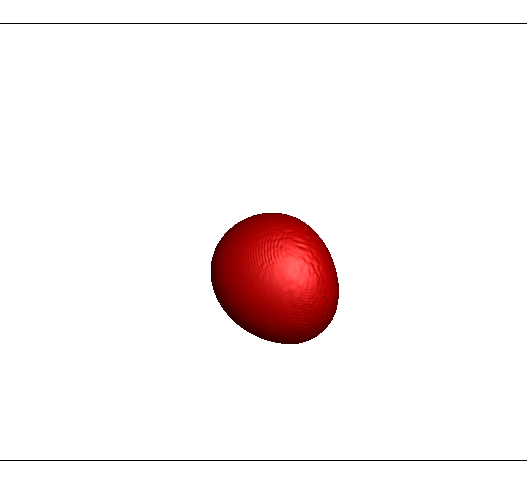}
    \caption{}
    \label{changing}
  \end{subfigure}
  \begin{subfigure}[b]{0.3\columnwidth}
    \includegraphics[width=\linewidth]{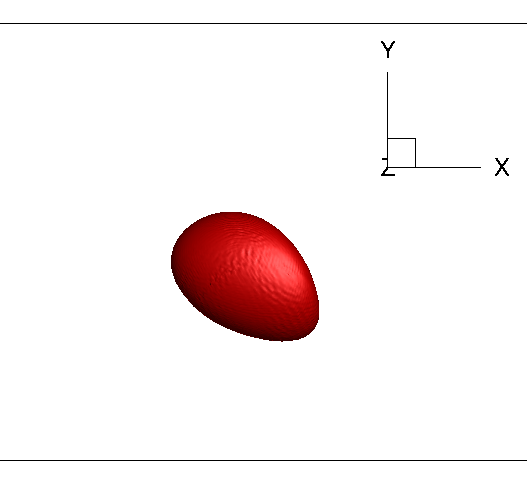}
    \caption{}
    \label{right to left}
  \end{subfigure}
  \begin{subfigure}[b]{0.3\columnwidth}
    \includegraphics[width=\linewidth]{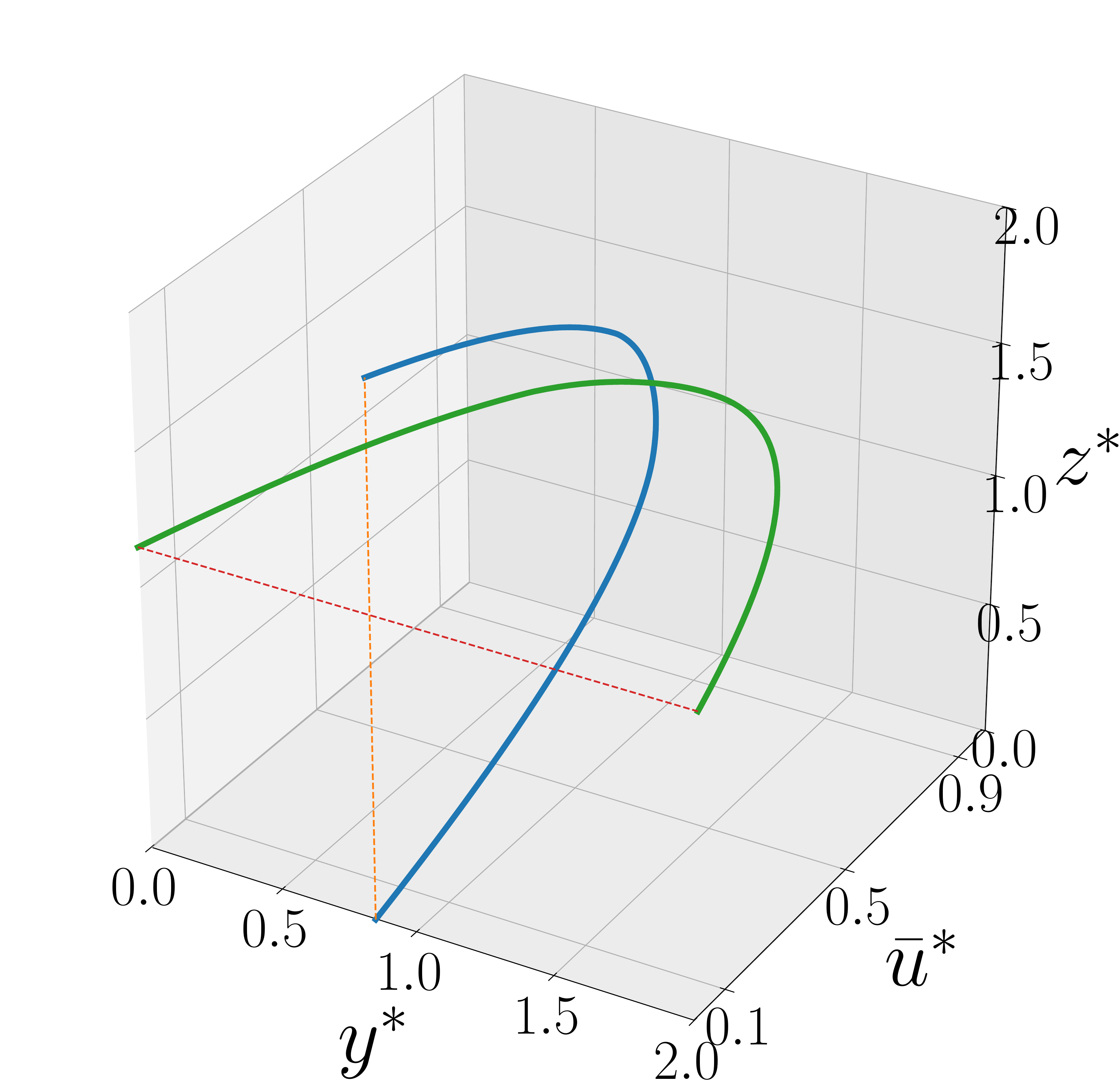}
    \caption{}
    \label{velocity left to right}
  \end{subfigure}
  \begin{subfigure}[b]{0.3\columnwidth}
    \includegraphics[width=\linewidth]{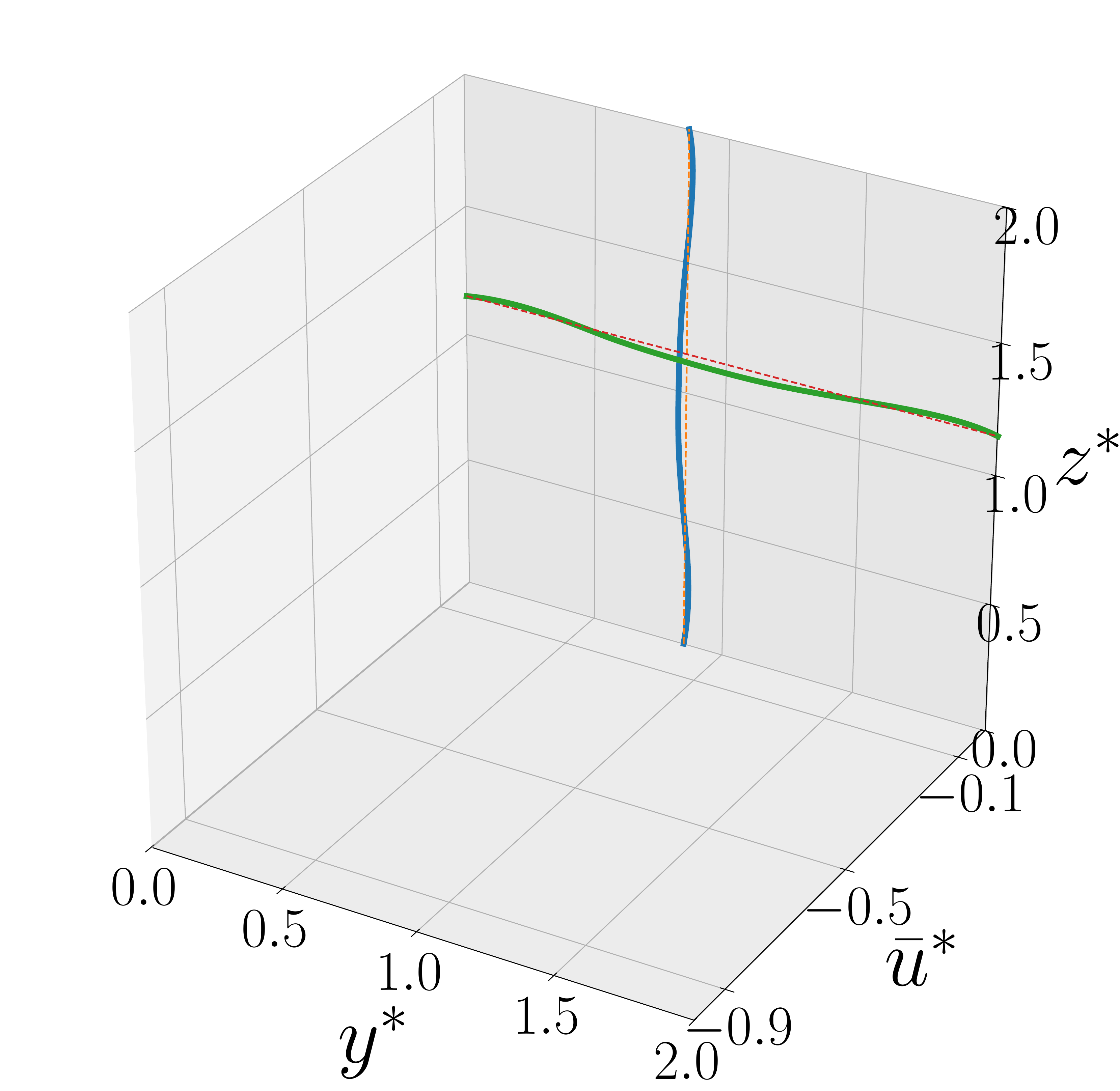}
    \caption{}
    \label{velocity changing}
  \end{subfigure}
  \begin{subfigure}[b]{0.3\columnwidth}
    \includegraphics[width=\linewidth]{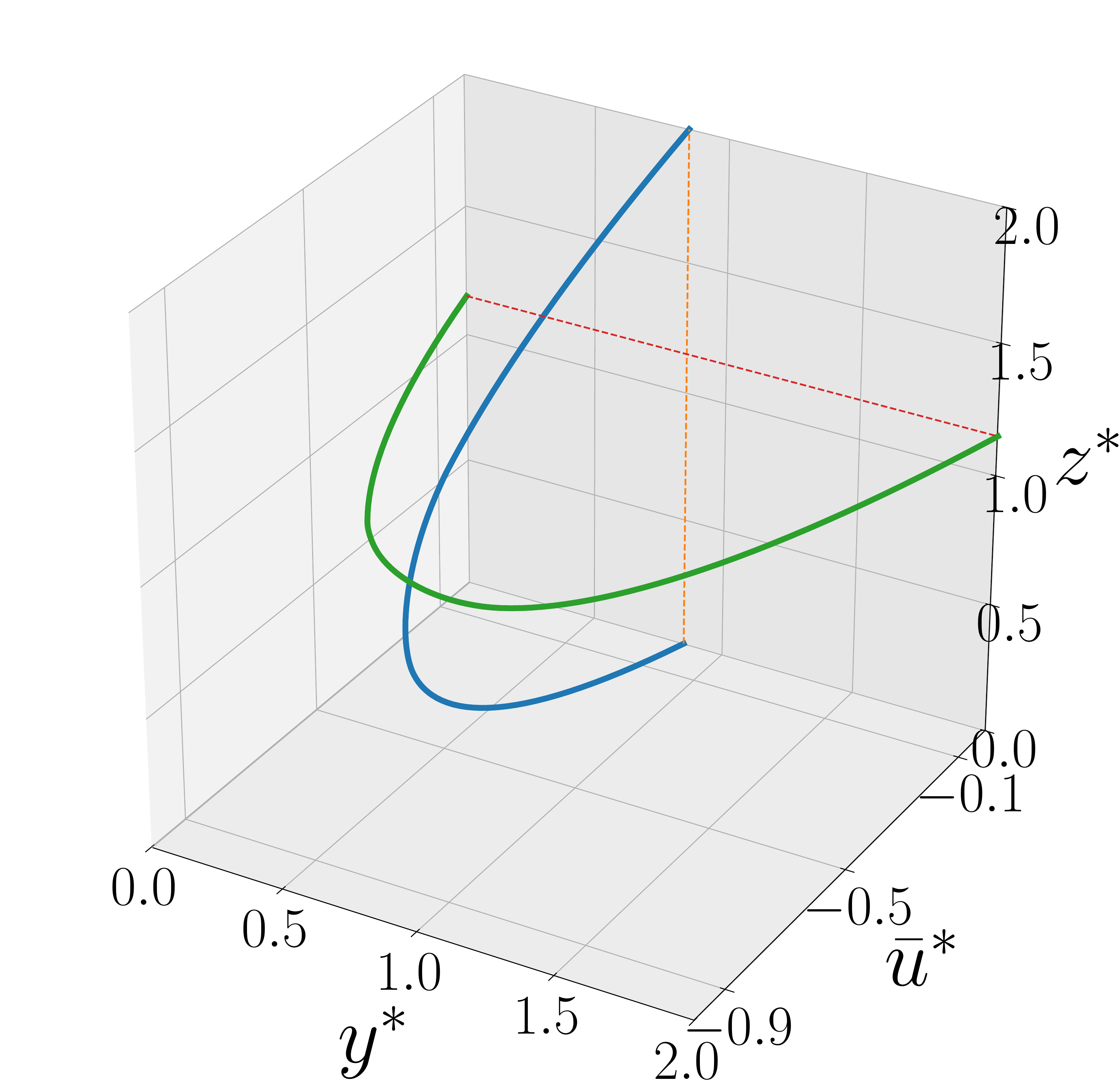}
    \caption{}
    \label{velocit right to left}
  \end{subfigure}
  \caption{Deformed shape of the capsule at $La=10$, $Re=37.8$, and $\omega^*=0.1$ in one cycle close to its focal point when flow is (a) from left to right, (b) changing direction, (c) from right to left, and the corresponding dimensionless averaged velocity profile along the flow direction when flow is (d) from left to right (e) changing direction (f) from right to left}
  \label{capsule shape}
\end{figure*}

\begin{figure}[h]
  \centering
  \includegraphics[width=3.7in,angle=0]{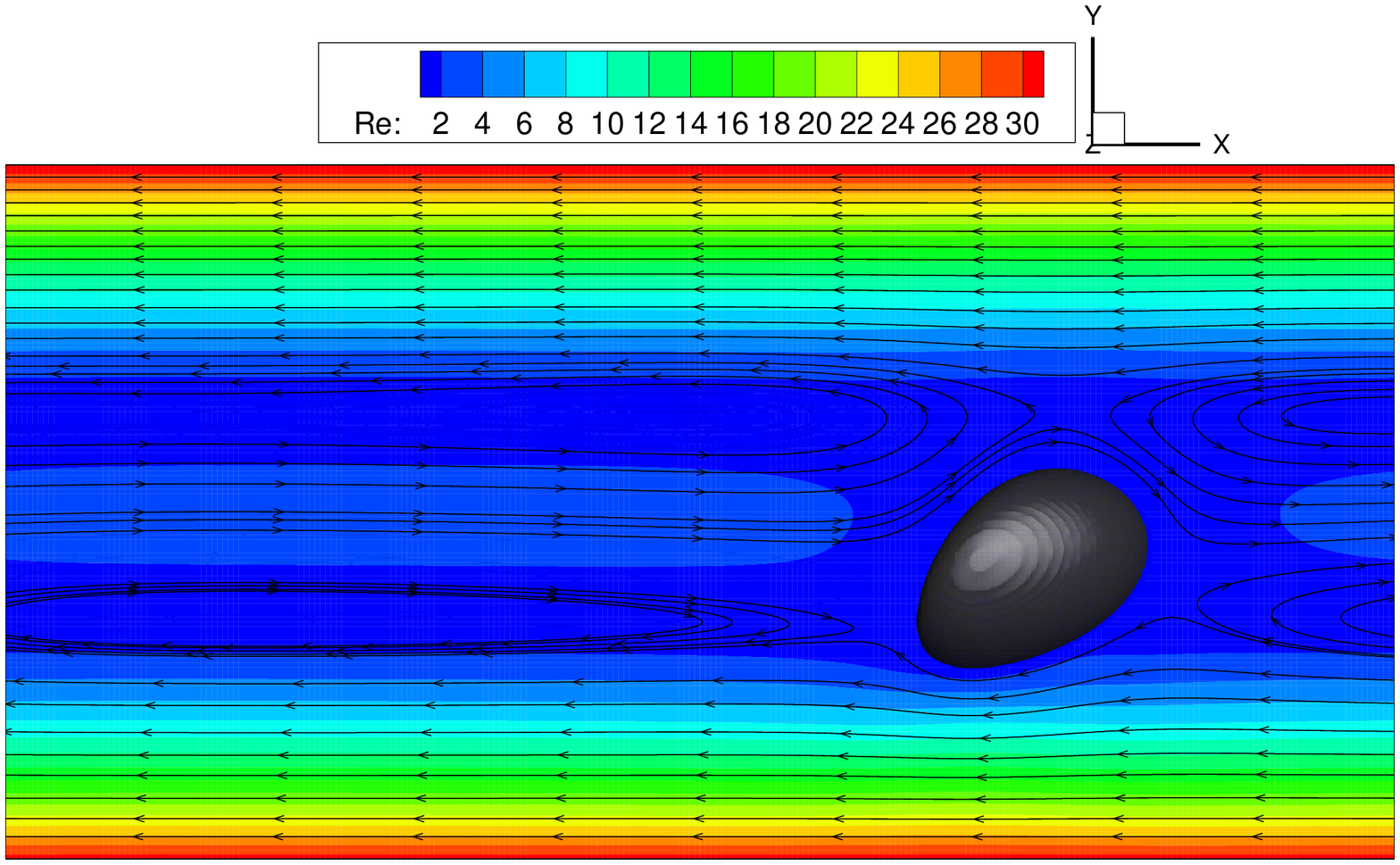}
   \caption{Flow field around the capsule on its frame of reference at $t^*=3269.87$ for $La=10$, $Re=37.8$, and $\omega^*=0.1$ on the $xy$ plane}
   \label{streamline}
\end{figure}

The Taylor deformation parameter of the capsule is defined as:
\begin{equation}
  D=\frac{L-B}{L+B}
\end{equation}
where $L$ is the principal major axis, and $B$ is the principal minor axis of an equivalent ellipsoidal particle. Fig. \ref{deformation} shows the oscillatory deformation of the capsule \cite{song2011dynamic, diaz2000transient} over time with a frequency twice the flow oscillation frequency since during each pressure gradient period, the capsule reaches maximum or minimum deformation twice \cite{zhu2015dynamics}. Fig.  \ref{capsule shape} shows the capsule at these instants of maximum and minimum deformations accompanied by the corresponding average flow velocity at the location of the capsule, and Fig. \ref {streamline} shows the surrounding flow field on the capsule frame of reference at the instant of maximum flow rate in one cycle near the capsule focal point. The minimum deformation occurs close to the instant when the flow is changing direction since the shear disappears. These two instants of time do not coincide exactly due to the present phase lag between the flow and capsule parameters (pressure gradient and capsule deformation in here) as discussed above. This shift in time increases by increasing the frequency due to the increase in the present hysteresis effect, as discussed earlier. Similarly, the maximum deformation occurs close to the instant of maximum flow rate, when the shear rate is also maximum. Fig. \ref{deformation} also confirms the directionality of the equilibrium point with respect to the frequency, as discussed previously. For the shown case in this figure, the average capsule deformation in each cycle reduces by increasing the frequency due to the reduction in the strength of shear rate \cite{zhao2011dynamics}. Furthermore, the average deformation decrease as the capsule approaches the center since the shear rate approaches zero, and the capsule surface energy gets closer to its minimum desired value. 

\subsection{\label{sec:level2}Necessary length for microchannel}
\subsubsection{\label{sec:level3}Effect of capsule deformability}

\begin{figure}[h]
  \centering
  \includegraphics[width=3.3in]{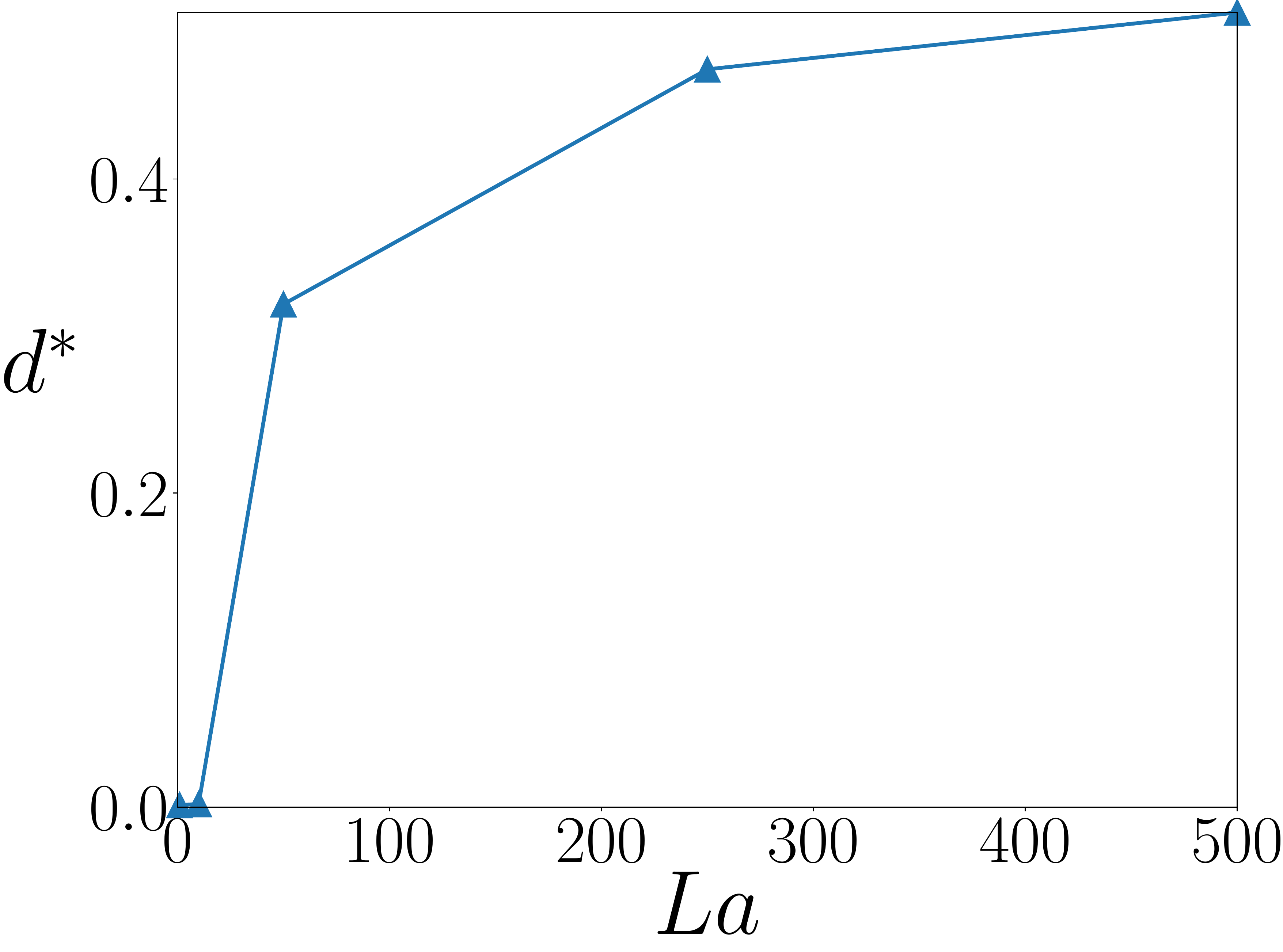}
   \caption{Distance of the capsule focusing position from the center at various $La$ numbers for $\omega^*=0.1$ and $Re=10$}
   \label{distance_La}
\end{figure}

Dynamics of capsules with different deformability have been studied in the channel. $La$ number quantifies the deformability, where a low $La$ denotes a very elastic capsule, whereas a high $La$ corresponds to a more rigid particle. Fig. \ref{distance_La} illustrates the distance of the capsule focusing position from the channel centerline for various $La$ numbers. Increasing the $La$ increases the distance, a behavior that is in good agreement with the previously published work by \citeauthor{raffiee2017elasto} \cite{raffiee2017elasto} This is because as the value of $La$ increases, the deformation-induced lift force, which is always towards the center \cite{fay2016cellular, zhang2016fundamentals} to minimize the surface energy of the capsule, reduces and therefore, the equilibrium position moves farther from the centerline. Furthermore, a relatively short microchannel can be designed for all the aforementioned capsules according to Table \ref{tbl:amplitude_La}. It can also be seen that increasing $La$ reduces the amplitude of movement, which is because as $La$ increases, capsules approach the wall where the velocity of the flow is lower than the centerline. Therefore, capsules with higher $La$ have lower average velocity and travel a shorter distance in the channel. This would be beneficial in particle focusing applications where the focusing length is very large due to small lift forces, e.g., small particles. It is worth mentioning that since the migration pattern of the capsule slightly depends on its initial location, a safety factor has to be considered for designing the microchannel based on the values reported in Table \ref{tbl:amplitude_La}.

\begin{table}[h]
\small
  \caption{Amplitude of capsule movement at various $La$ numbers for $\omega^*=0.1$ and $Re=10$}
  \label{tbl:amplitude_La}
  \begin{tabular*}{0.6\textwidth}{@{\extracolsep{\fill}}lll}
    \hline
    La & Amplitude of movement at the limit cycle \\
    \hline
    $1$ & $18.67$ \\
    $10$ & $18.65$ \\
    $50$ & $17.59$ \\
    $250$ & $15.89$ \\
    $500$ & $15.45$ \\
    \hline
  \end{tabular*}
\end{table}

\subsubsection{\label{sec:level3}Effect of channel flow rate}
\begin{figure}[h]
  \centering
  \includegraphics[width=3.3in]{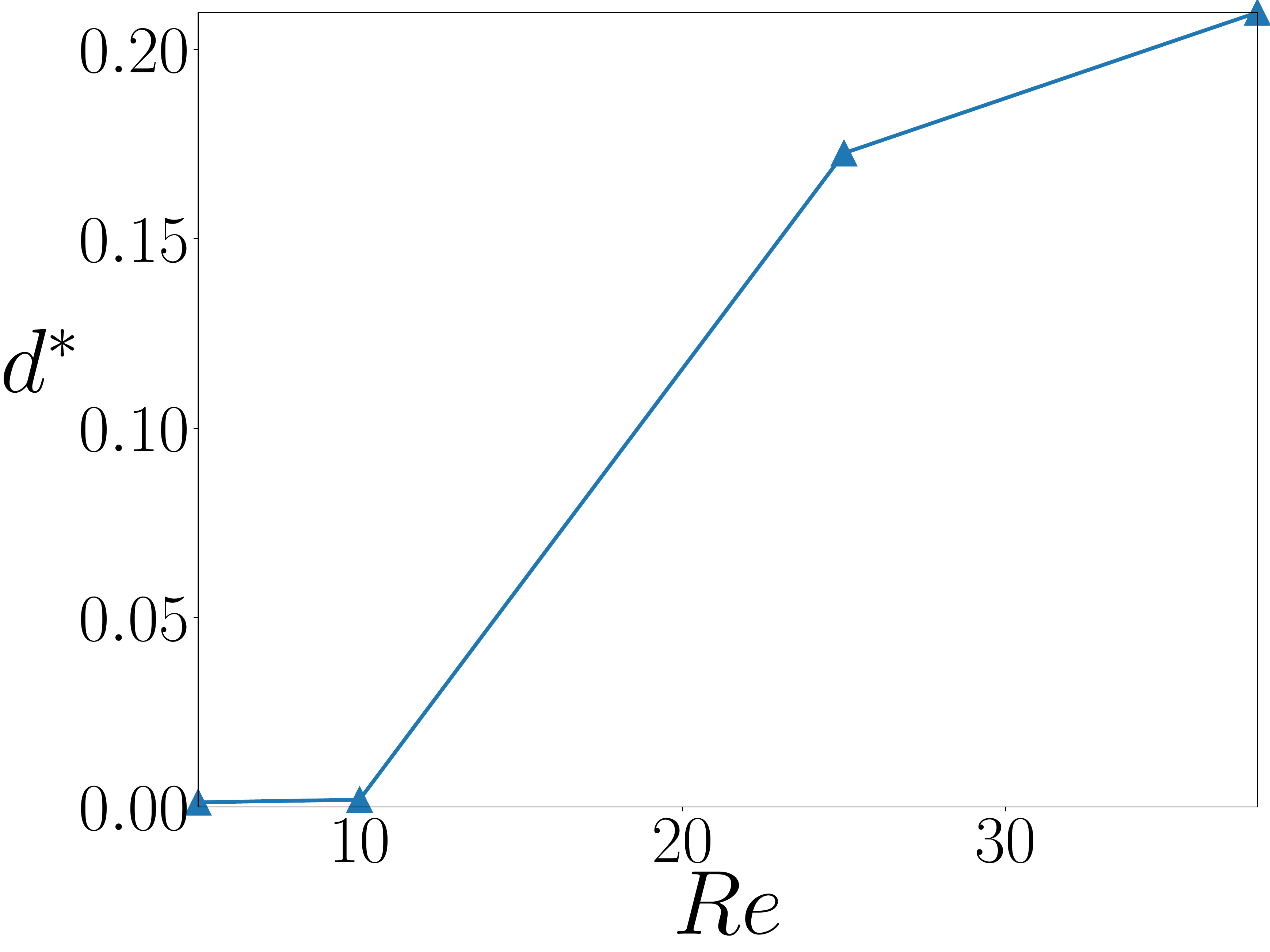}
   \caption{Distance of the capsule focusing position from the center at various $Re$ numbers for $\omega^*=0.1$ and $La=10$}
   \label{distance_Re}
\end{figure}

The effect of different flow rates in the channel on the capsule dynamics has been investigated in this section. Fig. \ref{distance_Re} illustrates the distance of the capsule focal position from the channel center as a function of Reynolds number. Increasing the $Re$ increases the distance from the center, which is in agreement with a previous work \cite{schaaf2017inertial}. This is because of the increase in the strength of the shear gradient induced force, acting towards the wall. Table \ref{tbl:amplitude_Re} shows a short necessary device length for all amounts of channel flow rates. It can further be noted that increasing $Re$ shortens the essential device length, which is because as $Re$ increases, capsules approach the wall where the velocity of the flow is lower than that of the center. It is vital to mention that the oscillatory flow does not produce a single focal point for capsule when $Re$ is very high ($Re=100$ for instance) as in these cases, the capsule oscillates drastically around a particular region rather than a single point due to the considerable variations in the values of both the shear gradient and deformation lift forces. This may force us to limit the system flow rate to some specific threshold to fulfill the purpose of interest.

\begin{table}[h]
\small
  \caption{Amplitude of capsule movement at various $Re$ numbers for $\omega^*=0.1$ and $La=10$}
  \label{tbl:amplitude_Re}
  \begin{tabular*}{0.6\textwidth}{@{\extracolsep{\fill}}lll}
    \hline
    Re & Amplitude of movement at the limit cycle\\
    \hline
    $5$ & $18.81$ \\
    $10$ & $18.65$ \\
    $25$ & $17.98$ \\
    $37.8$ & $17.17$ \\
    \hline
  \end{tabular*}
\end{table}

\subsubsection{\label{sec:level3}Square wave oscillatory flow}
\begin{figure}[h]
  \centering
  \includegraphics[width=3.3in]{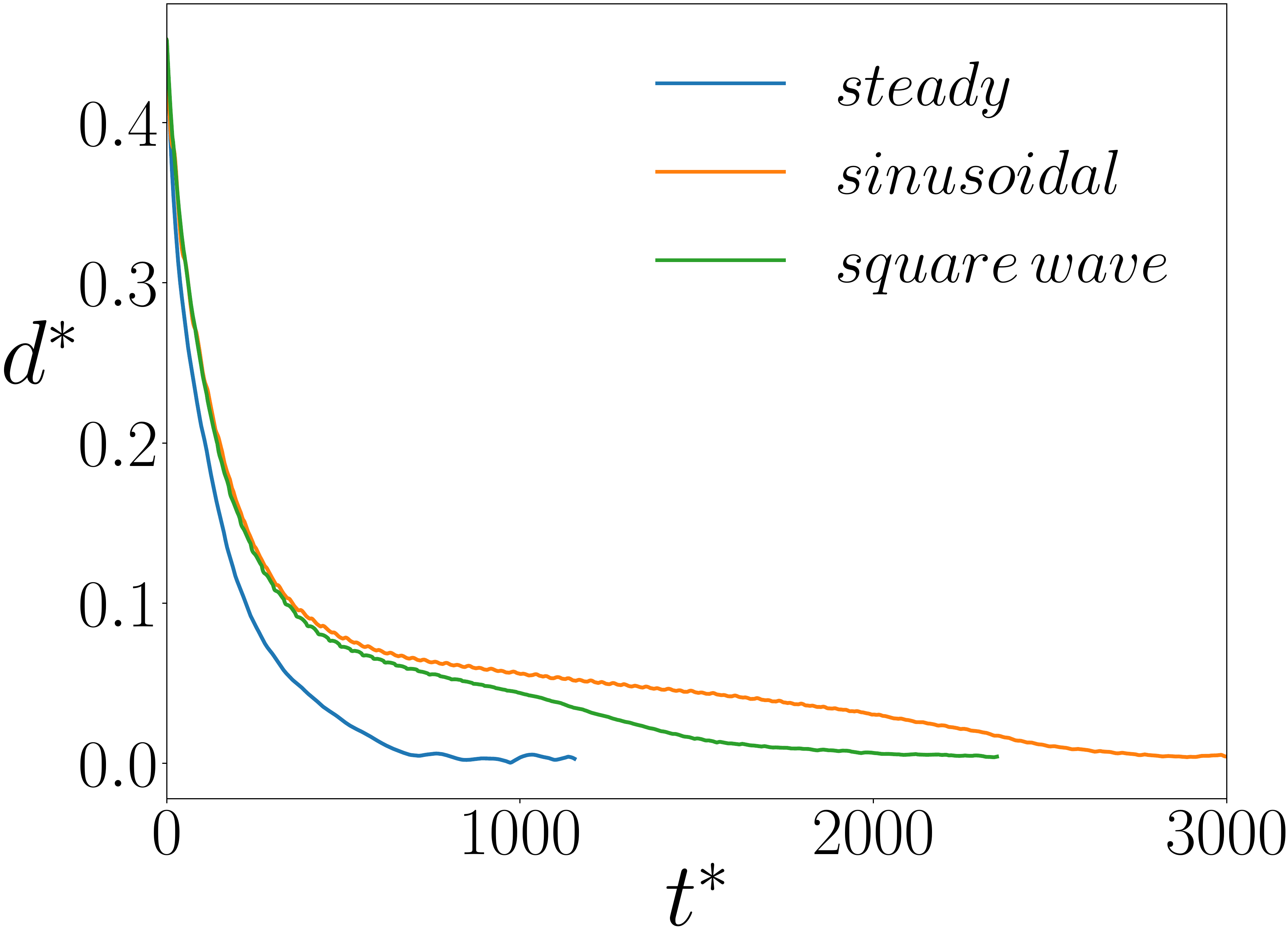}
   \caption{Evolution of the capsule motion for two types of oscillatory functions with $\omega^*=0.1$ and their comparison with the steady case at $La=10$ and $Re=10$}
   \label{fgr:square wave}
\end{figure}

In practice, it is easier to merely switch the direction of the flow in the channel by a control valve in the experiment. Thus, the simulations are also done with a pressure gradient having a square wave function with the same period and same area under the curve as that of the formerly defined cosine wave. This is to assure that the total amount of the external force acting on the fluid is the same for both cases. Hence, they both have the same average acceleration. The integral of a square wave is a triangular zigzag function, while integrating the cosine wave leads to a sine wave. Due to the shape of these two resultant functions, the zigzag has a sharp higher absolute peak compared to the smooth extremum of the sine, which results in a slightly higher average velocity because of which, the focusing time for the square wave function is less compared to that of the cosine wave. This is shown in Fig. \ref{fgr:square wave}. Furthermore, by plotting the hysteresis loops for both functions, it is found that the square wave has more hysteresis and requires a shorter device, which obeys the same pattern mentioned in the discussion of Fig. \ref{hysteresis}. This result is presented in Table \ref{tbl:square wave}.

\begin{table}[h]
\small
  \caption{Amplitude of capsule movement for two types of oscillation functions at $\omega^*=0.1$, $Re=10$, and $La=10$}
  \label{tbl:square wave}
  \begin{tabular*}{0.6\textwidth}{@{\extracolsep{\fill}}lll}
    \hline
    Function form & Amplitude of movement at the limit cycle \\
    \hline
    sinusoidal & $18.65$ \\
    square wave & $17.97$ \\
    \hline
  \end{tabular*}
\end{table}


\section{Conclusions}
We have studied the dynamics of a single deformable capsule in a rectangular microchannel with an oscillatory flow of a Newtonian fluid. We observed that adding sinusoidal and square wave oscillations change the focal position of the capsule compared to a steady flow case. At lower flow rates, an intermediate value of frequency signifies this change further for the same capsule deformability. At higher flow rates, however, increasing the frequency amplifies this change by moving the capsule equilibrium position towards the channel center. For the same frequency and flow rate, the change in the focal point is maximum at an intermediate capsule deformability. 
The oscillatory flow also enhances the focusing of micron-sized biological pathogens by significantly decreasing the essential microchannel length. 
This improvement is present for all types of capsules, flow rates, and any type of function or mechanism to produce the oscillatory behavior according to the cases presented in the current study. The higher the frequency, the lower is the essential length of the microfluidic device. The effect of varying $La$ and $Re$ numbers on the capsule dynamics was also studied. It was observed that higher flow rates and more rigid capsules shorten the required channel length. Generating the oscillatory flow using a square wave has the same effect as well. We believe that this work is significant because it provides the ability to have more direct control over the migration of cells inside the microchannels, which could be useful for cell sorting and separation, filtering bacteria and fungi, etc. 

\section*{Acknowledgement}
This work is supported by the National Science Foundation (Grant No. CBET-1705371) and by the USDA National Institute of Food and Agriculture (Hatch project 1017342).

\bibliography{sorsamp}

\end{document}